\documentclass[preprint,eqsecnum,aps]{revtex4}
\usepackage{graphicx}%
\usepackage{dcolumn}
\usepackage{amsmath}
\usepackage{psfig}
\makeatletter
\def\btt#1{\texttt{\@backslashchar#1}}%
\DeclareRobustCommand\bblash{\btt{\@backslashchar}}%
\makeatother

\begin{document}

\preprint{Co on Pt-Surf}

\title[Short Title]{Adsorption of CO on a Platinum (111) surface --
                   a study within a four-component relativistic density
                   functional approach}
\author{D.~Geschke, T.~Ba\c{s}tu\u{g}, T. Jacob, S. Fritzsche, 
        W.-D.~Sepp, B.~Fricke }
\affiliation{Fachbereich Physik, Universit\"at Kassel, D-34109 Kassel, Germany} 
\author{S.~Varga}
\affiliation{Department of Experimental Physics, Chalmers 
University of Technology, S-41296 G\"oteborg, Sweden}
\author{J. Anton}
\affiliation{Department of Atomic Physics, Stockholm University, Fescativ\"agen 24, S-10405
Stockholm, Sweden}

\date{\today}

\begin{abstract}
  We report on results of a theoretical study of the adsorption process of
  a single carbon oxide molecule on a Platinum (111) surface. A four-component
  relativistic density functional method was applied to account for a proper
  description of the strong relativistic effects. A limited number of atoms
  in the framework of a cluster approach is used to describe the surface.
  Different adsorption sites are investigated.
  We found that CO is preferably adsorbed at the top position.
\end{abstract}


\maketitle


\section{Introduction} \label{sec:introduction}

The interaction between atoms and surfaces is essential to understand
adsorption processes. In order to determine accurate structures,
vibration frequencies, electronic properties and forces {\it ab initio}
calculations are indispensable. 
Density functional theory (DFT) nowadays provides
a powerful tool in nearly all branches of fundamental research in 
microscopic regimes (see e.g. \cite{A93,BHRS99,NNP99,PY00}).
It also facilitates a detailed theoretical investigation of problems 
in surface physics and catalysis on a quantum-chemical level of theory
\cite{H98}, particularly the adsorption process of ad atoms on surfaces within  
the framework of a molecular orbital picture.
Due to the adsorption process on the surface the translation symmetry
parallel to the surface is violated. 
From a strict theoretical point of view methods of solid state physics are no  
longer applicable for this problem. 
In order to treat such systems approximately one can use two different approaches; 
one is the solid state method using super cells or a slab model, and the other is the molecular method
using clusters. As the size of the super cell or cluster increases, both methods should  
converge and yield the correct experimental value.  
In this paper we follow the molecular approach in which the physical system is 
approximated by a big molecule or cluster. 
Results of {\it ab initio} calculations are
comparable with experimental results for small molecular systems and/or
clusters.  
Such a cluster model is of course only an approximation for the 
description of unlimited surfaces and solids. Nevertheless, as long as 
the physical process which we are going to describe is strongly
localized, a cluster scheme may be an appropriate description. 
The adsorption process of a single atom or a simple molecule is
considered to be such a local effect.

One of the most extensively studied systems in this respect is the adsorption
of carbon oxide (CO) on a Platinum surface.
However, no good quantitative description has been achieved so far. 
The reason might be the lack of a proper description of the strong
relativistic effects resulting from the $5d$ band of Pt,
which lead to a different behaviour in the presence of Pt than 
on the surfaces of the homologue elements Ni and Pd \cite{PCKR97}. 

We have already tested this cluster approach on simpler systems successfully.
The adsorption of
a single Na and Ba atom at the Na(110) and the Ba(110) surface,
respectively, have been studied \cite{GFSFVA00}. The results were reasonable
and encouraged us to apply our method to the more complex system
widely used in catalytic processes: the adsorption of CO on a Pt(111)
surface.

In this paper we present the results of four-component relativistic
density functional calculations for the adsorption of CO on a Pt(111)
surface. Recently, our computer code has been developed to a new quality which
allows to calculate fragmentation energies and geometries of bigger clusters
with local and semi-local density functionals. Compared with scalar relativistic
approaches, the spin-orbit coupling is fully included in our formalism.

\section{What is known for the system CO at Pt(111) so far?}
  \label{history}

The adsorption of CO at a platinum surface seems to be one of the most
analyzed experimental systems. Apai et al. \cite{Apai} showed
that the CO molecule is adsorbed linearly with the C bonding towards the
surface which can be explained by the fact that the HOMO (5$\sigma$)
and LUMO (2$\pi^*$) of the CO are located at the C atom. So this site
is preferred for a binding with a surface.  The molecule does not
dissociate after adsorption and the main part of the binding is
dominated by the $d$ band of the platinum surface. The CO remains to
be effectively neutral.

A theoretical explanation of the binding formalism was given
by Blyholder \cite{Blyholder}. He studied the binding of CO at
metallic surfaces by using a H\"uckel method. Within this framework
the results show a formal charge transfer from the bonding 5$\sigma$
orbital of the CO towards the metal. On the other hand there is a
charge back transfer from the metal towards the anti bonding 2$\pi^*$ of
the molecule. This exchange of charge explains the fact that the
molecule is still neutral after adsorption.

With this theory one is able to distinguish the adsorption site at the
surface by the vibrational frequencies of the adsorbed molecule:
Effectively there is a charge transfer in the molecule from the
bonding 5$\sigma$ to the anti bonding $2\pi^*$ orbitals. The
occupation of the anti bonding orbital leads to a reduced electronic
charge between the C and O atom. This results in a weakening of the
bonding, and therefore the vibrational frequency of the adsorbed
molecule is lowered. The strength of this effect depends on the
adsorption site. A molecule adsorbed at a hollow position has a
stronger overlap between the metal band and the 2$\pi^*$ orbital than
at a top position. This is due to the fact that the molecule is closer
to the surface. Therefore the frequencies are lower at hollow
positions and one is able to distinguish the adsorption sites
\cite{Blyholder,Gantefoer}.

Ertl {\it et al.} \cite{Ertl} reported their results of the
adsorption of CO at the Pt(111) surface. They stated that the
strongest bonding arises at a hollow position which they erroneously
explained by the knowledge of similar systems like Ni and Pd. With
higher coverage the bridge position is occupied building a
$\sqrt(3)\times\sqrt(3)/R30^\circ$ overlay structure. In the regime of
zero coverage they estimated the binding energy to be $1.43 \pm 0.04
\,\,eV$. In the following years a lot of experimental papers appeared,
discussing the binding energy in this limit of low coverage. The
preferred adsorption site is now assumed to be the top position
followed by the bridge position
\cite{Froitzheim,McCabe,Horn,Hopster,Kelemen,Winicur,Steiniger,Seebauer}.
The results show an adsorption energy of $1.3$ to $1.6 \,\, eV$. All
these results have one thing in common: The adsorption energy is
calculated via the modified Clausius--Clapeyron formula of
thermodynamics: Pressure and temperature are varied whereas the
coverage is kept constant which is monitored by the
change of the work function. The resulting curve is a measure for the
adsorption energy. The adsorbed molecules form a dipole layer on the
surface resulting in a change of the work function which is
proportional to the coverage. In the regime of low coverage this
assumption may not be valid.

Newer results of Yeo {\it et al.} \cite{Yeo} show a significantly
stronger binding energy of $1.89\pm 0.20 \,\, eV$ at low coverage.
They used the new Single Crystal Adsorption Calorimetry (SCAC) method
which allows the direct measurement of the IR-photon emitted during
adsorption.

Somorjai {\it et al.} were able to estimate the binding distances of
the adsorbed molecules. They found a distance of $1.85 \pm 0.1$ {\AA}
between the C atom and the surface at the top position and a value of
$1.55 \pm 0.07$ {\AA} at the bridge position. The bridge position was
observed at a coverage of $\frac{1}{3}$. The binding distance within
the CO molecule after adsorption is measured to be $1.15 \pm 0.05$
{\AA} \cite{Ogletree,Blackman}. This is a small elongation of the
intra-molecular distance. The free CO molecule has a value of 
1.12 {\AA} \cite{Huber}.

On the theoretical side there are a lot of different approaches in the
estimation of the binding energy. Most of them use a cluster approach
\cite{Jennison,Ray,Gavezotti,Wong,Bala1,Bala2,Bala3,Ohnishi,Illas1,Illas2}
with different methods like modified H\"uckel methods or
pseudopotentials and different numbers of atoms. Some
\cite{Wong,Bala1,Bala2,Bala3} used only 1-2 atoms for the simulation
of the surface. The estimated binding energies at the top position are
between $0.19 \,\,eV$ and $4.21 \,\, eV$ reflecting the fact that
non-relativistic pseudo-potential or H\"uckel models are not able to give
reliable results for the adsorption energy. The main reason for this
seems to be the lack of reproducing a good behavior of the $d$
orbitals and therefore the $d$ band of Pt.  Models with less than 10
atoms in the cluster will probably only lead by chance to good
results.  The Atom Superposition and Electron Delocalization model
from Ray and Anderson \cite{Ray} based on extended H\"uckel method
combined with atomic one particle energies and atom--atom repulsion, on
the other hand, leads to acceptable results with a binding energy of
1.66 eV for the top, 1.26 eV for the bridge and 1.11 eV for the {\it
  fcc} hollow position followed by 1.03 eV for the {\it hcp} hollow
position. Thus the {\it fcc} hollow position is slightly preferred
among the both possible hollow positions.

Other authors used slab models \cite{Hammer,Philipsen2,KC99} or a super-cell
model \cite{Morikawa}.
However, all these methods were not applied in the regime of low coverage.
They work with
an effective coverage between $\frac{1}{4}$ and $\frac{1}{3}$. At
these coverages the lateral interactions of the adsorbed molecules are
not negligible. Yeo {\it et al.} \cite{Yeo} showed a decrease of the
energy by $0.2-0.4 \,\, eV$ within this regime. Thus the results they
get are in good agreement with experimental results but are not valid
at low coverage.

Brako {\it{et al.}} studied the on-top chemisorption of CO molecules as
a function of the coverage and the lateral relaxation of the platinum
surface using a classical Hamiltonian with an analytical model potential
\cite{BS98}.


\section{The theoretical method}\label{theorie}

We all know that an exact solution of the relativistic many--particle
Dirac equation is not possible. Large-scale multi-configuration or 
coupled cluster approaches are the best approximations for atoms so 
far. This high quality relativistic calculations are not yet possible 
for more than diatomic molecules.

For the size of systems we are discussing here only relativistic
pseudo-potential or density functional calculations are
available. The code which we are using \cite{relInteg} 
has its first roots in the relativistic version of the 
DVM approach \cite{roots}. 
Although the general structure of the density functional code has been
described in several publications we nevertheless repeat it here
in short because a number of details have been modified.

Following the Hohenberg--Kohn \cite{H.K.} and Kohn--Sham theorems 
\cite{K.S.} and starting from the  
no-pair approximation \cite{WDS} and also neglecting the minor important
contributions from spatial components of the four-current 
$j^v(\vec{r})=(j^0,\vec{j})$, the total energy (in atomic units) may be
written as 
\begin{equation} 
E[\rho]=T_s + E_{N}[\rho]+E_H[\rho]+E_{xc}[\rho] \label{DFT-functional}
\end{equation}
with the density 
\begin{equation}
j^0(\vec{r})=\rho(\vec{r})=
\sum_{-mc^2 < \varepsilon_k \le \varepsilon_F } 
\psi_k^{\dagger}(\vec{r})\psi_k(\vec{r})  \label{SCF-density}
\end{equation}
obtained from N one-particle Kohn--Sham Dirac spinors. 
This leads to the corresponding relativistic Kohn-Sham 
equations \cite{K.S.} 
\begin{eqnarray}
\left\{ T_s + V^N(\vec{r}) + V^C(\vec{r}) + 
V^{xc}(\vec{r})\right\} \, \psi_i(\vec{r})&=& \epsilon_i \, 
\psi_i(\vec{r}). \label{ET-Gleichungen}
\end{eqnarray}

Here $T_s=c\,\alpha_i \, p_i + (\beta -1 ) c^2$ is the relativistic
operator of the electronic kinetic energy, $V^N(\vec{r})$ represents
the nuclear potential assumed as a sum of point charges
\begin{equation}
V^N(\vec{r}) = \sum_\nu \frac{Z_{\nu e^2}}{\vec{R_{\nu}}-\vec{r}} \, ~,
\end{equation}
$V^{xc}(\vec{r})$ is the exchange--correlation potential derived from 
the exchange--correlation energy $E_{xc}$ 
\begin{equation}
V^{xc}(\vec{r}) = \frac{\delta E_{xc}[\rho]}{\delta \rho(\vec{r})} ~~.
\end{equation}

For the self-consistent solution of equation (\ref{ET-Gleichungen}) 
we use the relativistic local density approximation
\cite{Rajagopal,Rajagopal2} (RLDA) together with the Vosko, Wilk and Nusair
parameterization \cite{VWN} for correlation. Nonlocal corrections, 
with the relativistic form (RGGA) \cite{engelkellerdreizler} 
of Becke's 
GGA approximation\cite{B88} for exchange and Perdew functional of 
correlation \cite{P86} (B88/P86)is used perturbatively. 
Optionally we also have a relativistic extension of Perdew and Wang 
\cite{PW91} (PW91) for the xc-energy. 
%
%
The Coulomb potential $V^C(\vec{r})$ is given by 
\begin{equation} \label{Coulomb_exakt}
V^C(\vec{r}) = \int \frac{\rho(\vec{r'})}{|\vec{r}-\vec{r'}|}\, d^3r' .
\end{equation}
Numerical evaluation of the Coulomb matrix elements requires the
evaluation of the Coulomb potential at the grid points, which
leads to a large number ($n^2p$; $n$: number of basis functions, 
$p$: number of grid points) of nuclear attraction type integrals. 
It is possible to facilitate the $n^2p$ problem by approximating the 
exact density with an expansion in one center fit functions 
\begin{eqnarray}
  \widetilde\rho(\vec{r}) &=& \sum_{\alpha=1}^A\,\sum_{r=1}^{M_\alpha}\,
  \sum_{l=0}^{L_r}\, \sum_{m=-l}^l \, d_{rm}^{\alpha\, l} \, f_\alpha^r
  (\chi_\alpha) \, Y_l^m(\theta_\alpha,\phi_\alpha), \label{Modelldichte}
\end{eqnarray}
where $f_\alpha^r(\chi_\alpha)$ is the radial density of the
wave function $r$ of atom $\alpha$. $\chi_\alpha$ is a radial vector
centered at each atom $\vec{\chi_\alpha} = \vec{r} - \vec{R}_\alpha$.
$d_{rm}^{\alpha l}$ are the expansion coefficients of the approximate
density which are obtained via a least square fit to the density
(\ref{SCF-density}) 
\begin{eqnarray} \label{lsqf-den}
  \int (\rho(\vec{r})-\widetilde{\rho}(\vec{r}))^2\, d^3r  ~~ ,
\end{eqnarray}
or via minimizing the Coulomb energy of the difference density \cite{1staupaper}
\begin{eqnarray}
  \int \frac{(\rho(\vec{r})-\widetilde{\rho}(\vec{r}))
              (\rho(\vec{r})-\widetilde{\rho}(\vec{r}))}
	     {|\vec{r}-\vec{r'}|} \, d^3r' d^3r  ~~ .
\end{eqnarray}

For simpler
molecules, that means without complex density structure, it was
sufficient to use only the monopole part of the atomic wave functions
to build the approximate density. With the open $d$ shell of Pt it
is necessary to use higher moments in the expansion
(\ref{Modelldichte}). Actually we use values up to $l=2$.
Using this procedure the least square value of eqn. (\ref{lsqf-den}) 
was less than 5\%.

With the use of the approximate density (\ref{Modelldichte}) we can
reduce the three dimensional integral in equation
(\ref{Coulomb_exakt}) to one dimensional integrals:
\begin{eqnarray*}
  V(\vec{r}) & = & \sum_\alpha^A\,\sum_{r=1}^{M_\alpha}\,
  \sum_{l=0}^{L_r}\, \sum_{m=-l}^l \, d_{rm}^{\alpha l} \,V_{\alpha l }^{r m}(\chi_\alpha)
   \end{eqnarray*}
with
\begin{eqnarray*}
  V_{\alpha l }^{r m}(\chi_\alpha) &=&  \frac{4\, \pi}{2\,l+1}\, \frac{1 }{\xi^{l+1}_\alpha} 
                                        \,Y^m_l(\theta_\alpha,\phi_\alpha)\,
                                        \left[ \int\limits_0^{\chi_\alpha} {\xi'}_\alpha^{l+2} \, f_\alpha^r(\xi'_\alpha)
  d\xi'_\alpha + \xi^{2\,l +1}_\alpha \!\int\limits_{\chi_\alpha}^\infty \frac{1}{{\xi'}_\alpha^{l-1}} \,  f_\alpha^r(\xi'_\alpha) d\xi'_\alpha \right].
\end{eqnarray*}


In order to solve the Kohn-Sham equations (\ref{ET-Gleichungen}) we
use the Molecular-Orbital Linear Combination of
Atomic Orbitals (MO-LCAO)-method and expand the molecular orbital 
wave functions in  symmetry--adapted wave function $\chi_{j}$, which
themselves are expanded in atomic orbitals
$\chi_{n_{\nu}}({\vec{r}})$
which are atomic four--component Dirac--Spinors\cite{Meyer1,Meyer2}
\begin{eqnarray} \label{lcaoso}
\psi_i(\vec{r}) & = & \sum_j \chi_{j} (\vec{r}) c_{ji} \\
\chi_{j} (\vec{r}) & = & \sum_{n_{\nu}} \chi_{n_{\nu}}(\vec{r})
d_{n_{\nu}j} 
\label{AO-Ansatz}
\end{eqnarray}
with $n_{\nu}=(\nu,n,\kappa,m)$. Here 
$\chi_{n_{\nu}}(\vec{r})$ are the relativistic four--component 
atomic basis functions centered at the atoms. Using this symmetry 
adapted atomic basis set implies the non pair approximation \cite{WDS}.
The symmetry coefficients $d_{n_{\nu}j}$ are calculated by the program 
TSYM of Meyer \cite{Meyer1,Meyer2}. The expansion coefficients 
$c_{ij}$ of the molecular wave functions together with the one-particle 
energies $\epsilon_i$ are the
results of the SCF calculation of equation (\ref{ET-Gleichungen}).
Inserting (\ref{lcaoso}) into (\ref{ET-Gleichungen}) gives the matrix
secular equation in the symmetry orbital basis $\chi$,
\begin{equation}
{\bf H} {\bf C} = \varepsilon {\bf S} {\bf C} \label{secular}
\end{equation}
where {\bf H} and {\bf S} are the Hamiltonian and overlap matrices respectively,
{\bf C} is the coefficient-matrix ($c_{ij}$) and $\varepsilon$ is 
the eigenvalue-vector ($\varepsilon_i$).
The matrix elements, $H_{ij}$ and $S_{ij}$ are evaluated
numerically using the multi-center integration scheme by Baerends et
al. \cite{Baerends,Baerends2}. In this scheme the whole molecular volume is 
spatially partitioned into so called Voronoi cells and each cell is 
divided into an atomic sphere around the atomic nucleus and polyhedra.
This method was extended by a special transformation to account for 
the relativistic behavior of the wave function near the nuclei  
\cite{relInteg}.

\section{The Frozen--Core approximation}
In order to reduce the computational requirement one very popular way 
is to use either pseudopotentials or the Frozen Core approximation 
(FCA).
Both approximations are based on the idea that chemical binding 
is realized via the outer valence orbitals only. 
In our calculations we used the Frozen Core Approximation.
In FCA it is assumed that the inner orbitals do not change their form 
during the adsorption process so that they can be kept fixed
in the self consistence procedure. 
This reduces the numerical effort for the heavier elements like Pt 
drastically.  

Since the FCA is a very popular approximation and used in almost
all kinds of {\it ab initio} calculations (from density functional 
to quantum chemical methods) there are many references in the 
literature \cite{FC1,FC2,FC3} for this method. 

Although the principal procedure of the FCA in all {\it ab initio} 
methods is the same, there are 
differences in practical implementations.
Therefore we give a short summary of the FCA in our code.
Compared with all electron calculations there are two 
differences in FC calculations: the orthogonalization of the valence 
orbitals to the core orbitals and the calculation of the total energy.

The FCA is based on a decomposition of the Hilbert space into 
a core $(c)$ and a valence $(v)$ part. Accordingly, the molecular 
wave functions can be written as: 
\[ 
|\psi\rangle = |\psi_v \rangle + |\psi_c \rangle .
\] 
In order to ensure the constrained Ritz variation of the valence 
orbitals, a core-valence orthogonalization has to be carried out. 
In our implementation we followed the idea of the normalization of 
the overlap matrix in the core-valence orthogonalization. 
The decomposition of the Hilbert space results in the 
decomposition of the overlap and Fock matrices into core-core, 
core-valence and valence-valence parts which have to be orthogonalized 
to each other. 

In order to ensure the core-valence orthogonalization 
all contributions of core 
orbitals are removed from the valence orbitals. This is achieved 
by the explicit orthogonalization of the valence orbitals onto the 
core orbitals. 
\begin{eqnarray*}
  |\chi_v'\rangle &=& |\chi_v\rangle - S_{vc}\, S_{cc}^{-1}\,
   |\chi_c\rangle .
\end{eqnarray*}
Here $S_{vc}$ is the valence--core overlap matrix. 
As a result the transformed overlap and Fock matrices of the
whole system are now of the form:
\begin{eqnarray} \label{fc-SF}
 \mbox{S'}= \left( {S_{cc} \atop 0} \, { 0 \atop S'_{vv}} \right) 
     ~~~;~~~ 
  \underline{\underline{\mbox{H}}} = 
             \left({H_{cc} \atop 0}\,{0\atop H'_{vv}}\right) +
 \underbrace{\left({ 0 \atop H'_{cv}} \, {H'_{vc} \atop 0} \right)}_{
               \rightarrow neglected}.
\end{eqnarray}
We assume that the core orbitals of different atoms do not overlap. 
By the use of atomic orbitals the core orbitals of the same atom are 
by default orthogonal. Therefore the corresponding symmetry orbitals 
are also orthonormal and $S_{cc}=1$.

Due to the chemical binding the relative change in the core 
potential is small. The change in the core wave functions which is
a second order effect can therefore be neglected. 
This means that $H'_{vc}$ in eqn. (\ref{fc-SF}) can be neglected. 
As a result of this approximation the secular equation (\ref{secular})
decouples into two equations, which are solved like all electron 
calculations. 
Both approximations define our implementation of FCA.

The total energy within this Frozen--Core approximation is now:
\begin{eqnarray}
E &=&  \sum_{\alpha=1}^A\sum_{\mu=1}^{N^c_\alpha} n^c_{\alpha \mu} \,
\epsilon_\mu^\alpha - \frac{1}{2}\,\sum_{\alpha=1}^A \int v_{c \alpha}^C \,
\rho_c^\alpha \, d^3r \nonumber \\
&& - \, \sum_{\alpha=1}^A \int v_{v \alpha}^C \, \rho_c^\alpha
\, d^3r - \sum_{\alpha=1}^A \int v_\alpha^{xc} \, \rho_c^\alpha\, d^3r\nonumber\\
&& + \, \sum_{\alpha=1 }^A \sum_{\beta\neq \alpha }^A \int v_\alpha^N \, \rho_c^\beta \,
d^3r + \frac{1}{2} \sum_{\alpha=1}^A 
\sum_{\beta\neq \alpha }^A \int v_{c\alpha}^C \, \rho_c^\beta \, d^3r \nonumber \\
&& +\, \sum_{i=M^c+1}^{M} n_i^v \, \epsilon_i -\frac{1}{2} \int \widetilde{\rho}^v \,
\widetilde{V}^C_v \, d^3r -\int V^{xc} \rho_v \, d^3r \nonumber\\
&&+ E^{xc} + \sum_{\alpha,
  \beta =1 \atop \alpha < \beta}^A \frac{Z_\alpha \,
  Z_\beta}{|\vec{R}_\alpha-\vec{R}_\beta|}. \label{TotaleFCEnergie}
\end{eqnarray}
The symbols with a small $v$ represent atomic potentials. The subindices $c$
and $v$ mean that the core and valence densities are used to calculate the
potentials. The first two lines of equation (\ref{TotaleFCEnergie}) consist of
pure atomic contributions. As far as we do not change the basis functions
these values can be neglected as long as we are only interested in binding energies
and distances. In this case these core contributions cancel completely
out. The third line describes only the core--core interactions of different
atoms. These values can be approximated by the use of point charges:
\begin{eqnarray*}
  \sum_{\alpha, \beta=1 \atop \beta\neq \alpha }^A \int v_\alpha^N \,
  \rho_c^\beta \, 
d^3r + \frac{1}{2} \sum_{\alpha, \beta=1 \atop \beta \neq \alpha}^A \int v_{c
  \alpha}^C \, \rho_c^\beta \, d^3r &=&  \sum_{\alpha=1 }^A \sum_{\beta\neq
  \alpha }^A \int \rho_c^\beta \, (v_\alpha^N +v_{c \alpha}^C)  \,
d^3r\\
&\approx&  -\,\sum_{\alpha=1 }^A \sum_{\beta\neq \alpha }^A \frac{Z^c_\beta
   \,  (Z_\alpha-\frac{1}{2}Z^c_\alpha) }{|\vec{R}_\alpha-\vec{R}_\beta |}.
\end{eqnarray*}
The last two lines are exactly the same as in the total energy expression 
in an all electron calculation 
but only with the use of the new valence orbitals for the density. For
$V^{xc}$ we still need the full density $\rho = \rho_c + \rho_v$.

\section{Results}

Our aim in this paper is the {\it ab initio} calculation of the
adsorption of CO at Pt(111) surface. This process is one of the most
frequently used catalytic processes but is not understood in detail up 
to now. We have referred to the experimental findings and theoretical
methods which were applied up to now in section \ref{history}.

As Pt is a very heavy metal where relativistic effects play
an important role we at least have the appropriate
method. Unfortunately the computational effort restricts the actual
calculations to a very limited number of Pt atoms which can be used in
a real calculation. Nevertheless we have achieved to converge clusters
between 7 and 13 Pt atoms depending on the adsorption site of the CO
molecule. To keep the calculation time in an acceptable range, we in
addition used only a quasi minimal basis set which was optimized via
the degree of ionization of the atomic wave functions used as basis
functions. For Pt we use the $1s$ to $6p$ wave functions
with a degree of ionization of $+0.2$. As an effect this leads to
valence functions which are slightly more localized than the neutral
ones but are more suitable to reproduce the binding character of the
bulk. For the wave functions with the main quantum numbers n=1 to 4 
in addition the FCA was used.

\subsection{The CO molecule}

For the calculation of the CO molecule a larger basis set was used.
This system requires a very good basis to account for the strong 
chemical bonding and charge distribution within this molecule.

The best results for the CO molecule were obtained with a minimal 
basis set consisting of the $1s-2p$ wave functions of the neutral 
atoms in combination with additional $3p,3d$ wave functions with 
an degree of ionization of +4.0. In this molecule all electrons were 
used in the SCF process.

Table \ref{Multipoltabelle} shows the results for the CO molecule, 
which are close to the experimental data.  The negative value of 
the dipole moment indicates a major negative charge at the C atom. 
The binding energy is slightly too high, which may be attributed to 
an unpolarized treatment because current-dependent functionals are 
not available.

\subsection{The adsorption calculations}

In order to calculate the adsorption energy of CO on the Pt(111)
surface one first has to define the clusters which
simulate the surface for the four possible adsorption sites of the CO
molecule. The internuclear axis of the CO molecule is assumed to be
perpendicular to the surface with C towards the surface.
The four possible adsorption sites are the top,
bridge, fcc--hollow and hcp--hollow of the (111) surface of Pt.
The different adsorption sites are shown in Figures \ref{Top_multipol}
to \ref{Hcp_multipol}.
As insert in the figures \ref{Top_multipol} to \ref{Hcp_multipol} we
present the Pt--clusters as molecular models in the view from above
and in a perspective view for the four adsorption sites. The CO
molecule can also be seen. The distances between the Pt atoms are kept
fixed to their bulk values ($a_0=3.92$ \AA) and the CO distance to
the value calculated above.

In the cases of top and the hollow positions the $C_{3v}$ symmetry was utilized.
For the bridge position only a $C_{1h}$ symmetry can be used.
Figures \ref{Top_multipol} to \ref{Hcp_multipol} show the results of
the calculations. Because of the low symmetry of the bridge position
a maximum of 7 atoms had to be used to get results in a realistic 
calculation time.
The notation Pt$_{m+n}(m,n)$ denotes the number of
atoms per layer: $m$ atoms in the first layer and $n$ in the second
resulting in $m+n$ total Pt atoms. The bulk was represented only by two
layers. As shown in the literature, this number of layers seems to be
necessary and nearly sufficient \cite{Wiesenekker,Philipsen,Castells}.

Results of our calculations for binding of the CO
molecule on the four surfaces are given in Tables
\ref{Tabelle_mult1}--\ref{Vergleich-Abstand} for the GGA calculations
according to Becke 1988 \cite{B88} in combination with Perdew 1986
\cite{P86} (which in short term notation is B88/P86).

The results show the experimentally observed binding order with the top
position as the strongest one and therefore as the preferred 
adsorption site. Like Ray and Anderson \cite{Ray} we also find a 
slight preference of the fcc
position in the hollow positions.

Within our method the binding energies are by about $10 - 20
\%$ too high compared with experimental results (see for instance the
results of the CO molecule). Having this in mind our results of 2.3
$eV$ binding energy at the top position compares well
with the measured value of $1.86 \pm 0.20$ $eV$ of Yeo {\it et al.} 
\cite{Yeo} (see table \ref{Tabelle_mult1}).
A direct comparison of binding energies at other positions with experimental
results is not available. The calculations here were performed with a
single CO molecule only. In experiments an occupation of the other
adsorption sites is only observed at much higher coverages which are not considered
here.

In Table \ref{Tabelle_mult2} the calculated bond distances of the C
atom (of the CO molecule) with respect to the surface at the energy 
minimum at the four
possible adsorption sites are listed.

In addition very informative is Table \ref{Vergleich-Abstand}. Here
the distances are calculated towards the next Pt atom of the surface
for all four CO positions.  For better comparison with the
experimental results these values are converted to \AA. These results
correlate with the binding behaviour.  The distance at top position is
the shortest whereas the hollow positions have the greatest distances.
A comparison with the experimental values shows a good agreement for the top
position as well as the bridge position where our value is slightly
shorter than the experimental value. One reason
for this may lie in the effective coverage when the bridge
position is experimentally observed. As already mentioned, 
the interactions of the adsorbed CO molecules are not negligible
\cite{Jennison} within this coverage.

Table \ref{theory-comparision} presents a comparison of our 
values for the top--position with other theoretical results. 
These values demonstrate that the bond energies still do scatter 
up to 1 eV. The same applies for the various theoretical 
results for the hollow--position in Table \ref{theory-comparision1}.

Table \ref{Mulliken_CO} shows the electronic charge distribution of 
the CO molecule as a result of a Mulliken population analysis of our 
calculations. The charge of the oxygen atom remains more or less 
constant and independent of the adsorption process and adsorption 
site whereas the carbon atom changes its charge by small amounts. 
A more detailed analysis of the electronic charge distribution 
shows that the charge between the two atoms C and O is
reduced in the direction off the molecular axis.

It is further interesting to see that the CO molecule as a whole has
almost no charge transfer at the top position which has the strongest
binding whereas all three other positions show a negative charge
transfer from the cluster to the molecule of nearly half an electron.

In addition we studied the effect of elongation of the CO
distance on the top--position adsorption site. As a result we 
found only a very small elongation of $0.01-0.02\,\,a.u.$ compared with
the free molecule.

\section{Conclusion}

Using a cluster approach within a relativistic density functional 
approximation as described above, we get results for the adsorption 
of the CO molecule on the Pt(111) surface which are reasonably
good. This statement in itself is very remarkable although we must
admit that we are not able to show the convergence of these results 
with cluster size here. For one of the positions we have been able
to include more Pt atoms in the cluster and in fact, the binding 
energy did not change any more.
As expected, the binding energy is slightly higher than the
experimental value of Yeo {\it et al.} \cite{Yeo}.

Finally we note that the used method is able to describe the
adsorption process. We are able to calculate this important system
with a quasi {\it ab initio} method (the only experimental result used
in the calculation is the structure of the bulk) in a fully relativistic
framework. In our calculations all relativistic $d$ orbitals of Pt are
treated dynamically.

\section{Acknowledgment}
T.B. gratefully acknowledges support from the Alexander von Humboldt
Foundation (AvH), T.J. from the Studienstiftung des Deutschen Volkes,
S.V. from the German Academic Exchange Service (DAAD) and J.A. 
from the Deutsche Forschungsgemeinschaft.

\newpage
\section*{Table Capture}
\begin{description}
\item [Table \ref{Multipoltabelle}.]
Results for the CO molecule: binding energy (BE), binding distance 
(d$_{C-O}$), ionisation potential (IP), frequency ($\omega_{C-O}$) 
of the  molecular stretch vibration and dipole moment ($\mu$). The 
dipole moment was calculated at the distance of the energy minimum. 
Assuming the experimental binding distance a value of $\mu = -0.052$ 
a.u. was obtained.
\item [Table \ref{Tabelle_mult1}.]
Results for the binding energy of CO on Pt(111) in eV for the four 
adsorption sites.
\item [Table \ref{Tabelle_mult2}.]
Results for the distance of the carbon atom to the surface (in a.u.) 
for the four bond positions.
\item [Table \ref{Vergleich-Abstand}.]
Comparison of the results of the distance of the carbon atom to the 
next surface platinum atom (in a.u.).
\item [Table \ref{theory-comparision}.]
Comparison of theoretical results for the top--position.
\item [Table \ref{theory-comparision1}.]
Comparison of theoretical results for the hollow--position.
\item [Table \ref{Mulliken_CO}.]
Electronic charge of the adsorbed molecule and the consisting atoms as 
a result of a Mulliken population analysis for CO at Pt(111).
\end{description}
\newpage
\section*{Figure Capture}

\begin{figure}[t]
\centerline{\psfig{figure=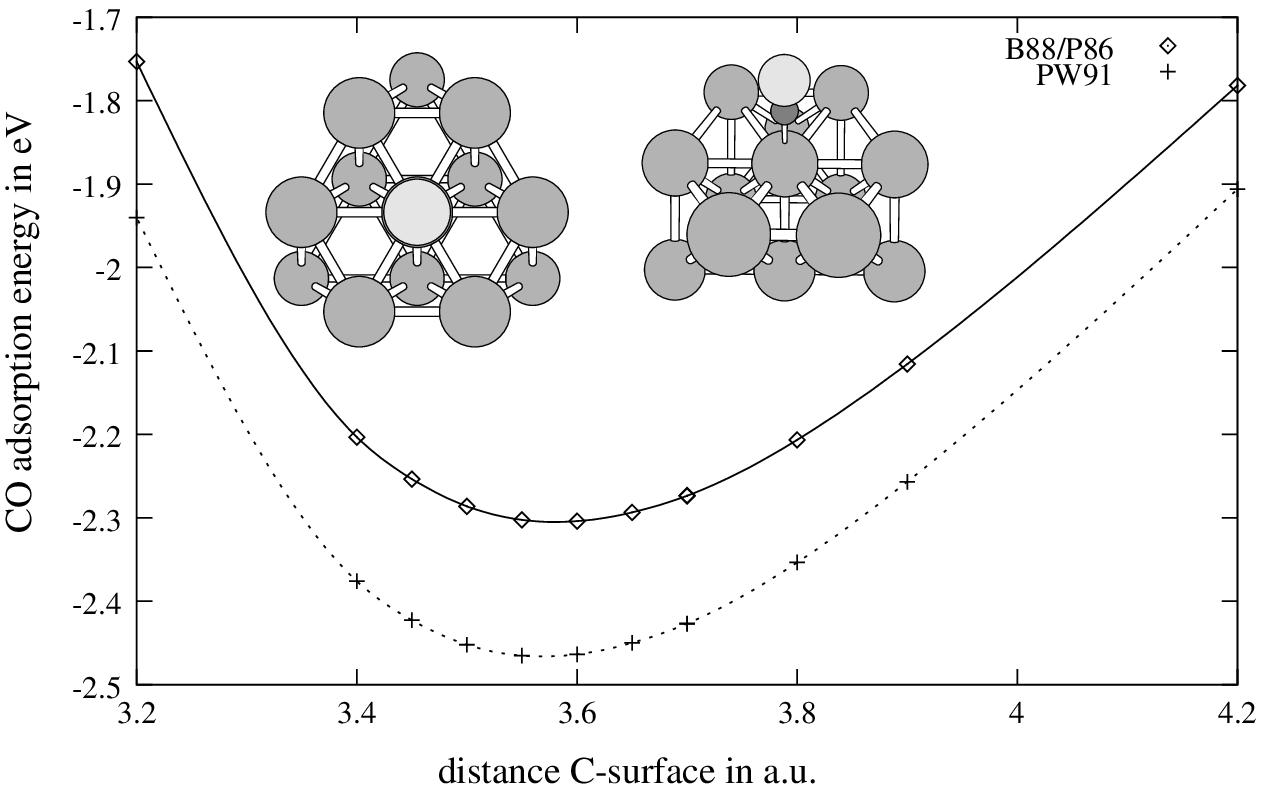,width=13cm,height=8cm}}
\caption{}\label{Top_multipol}
\end{figure}

\begin{figure}[p]
\centerline{\psfig{figure=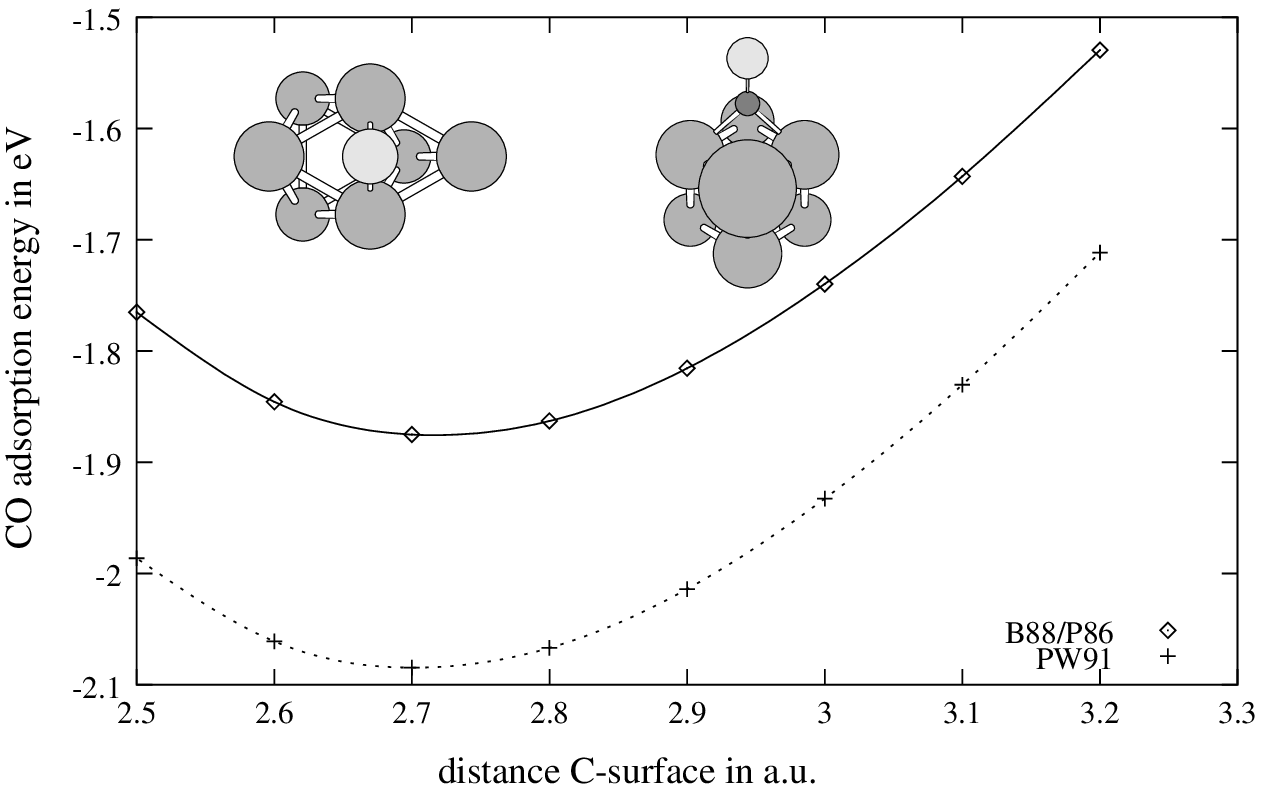,width=13cm,height=8cm}}
\caption{} \label{Bridge_multipol}
\end{figure}

\begin{figure}[p]
\centerline{\psfig{figure=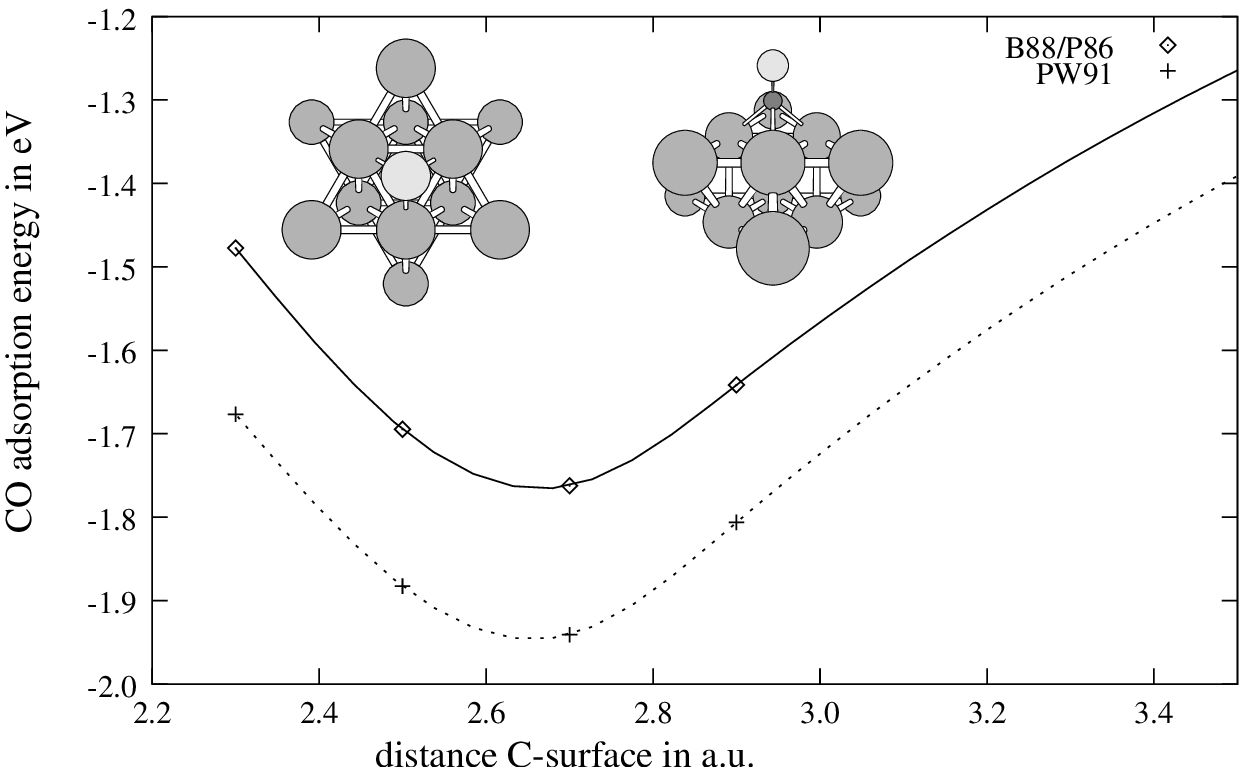,width=13cm,height=8cm}}
\caption{} \label{Fcc_multipol}
\end{figure}

\begin{figure}[p]
\centerline{\psfig{figure=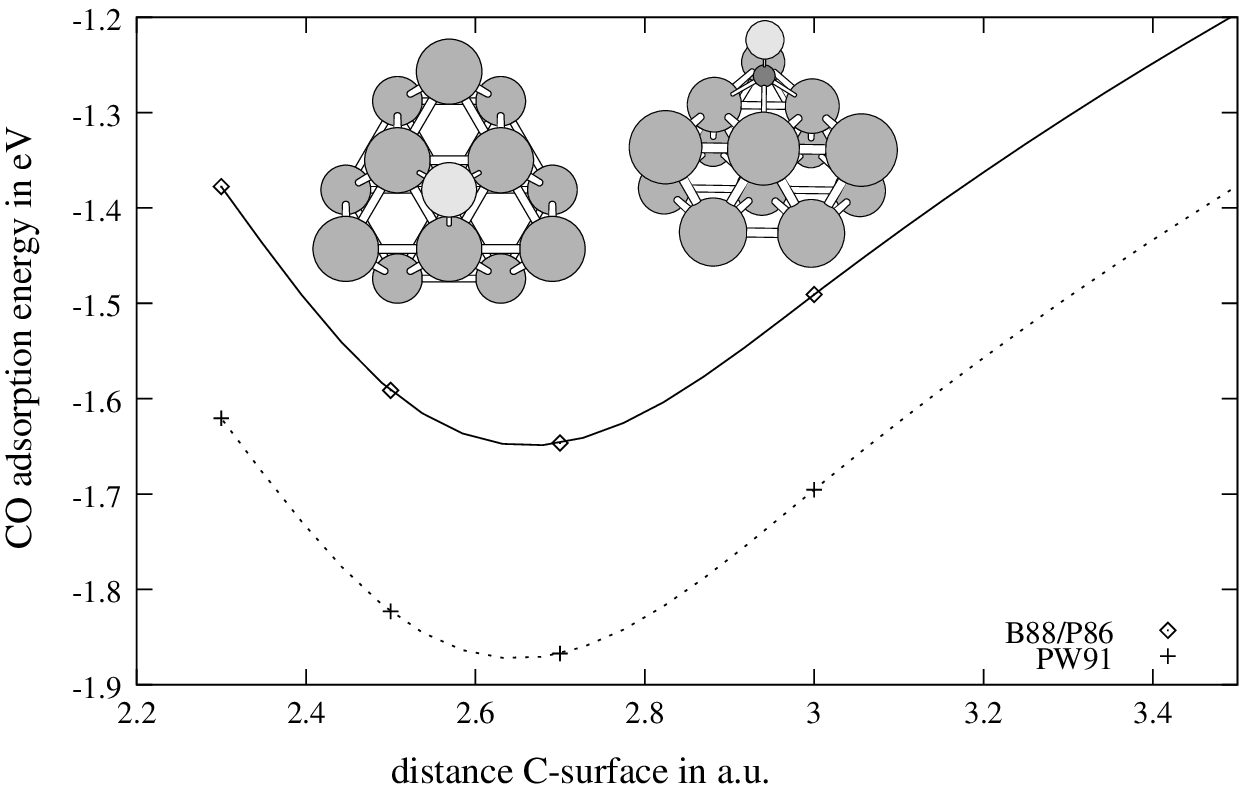,width=13cm,height=8cm}}
\caption{} \label{Hcp_multipol}
\end{figure}

\begin{description}
\item [Figure \ref{Top_multipol}.] Potential energy curve of CO on Pt(111) at 
                                   the top position. The platinum cluster 
				   consists of 13 atoms: Pt$_{13}(7,6)$.
				   Solid curve is obtained with B88
				   exchange and BP86 correlation potential
				   and dotted curve is obtained with PW91
				   exchange and correlation potential.
\item [Figure \ref{Bridge_multipol}.] Potential energy curve of CO on Pt(111) 
                                      at the bridge position. The platinum 
				      cluster onsists of 7 atoms: Pt$_{7}(4,3)$.
				      Solid curve is obtained with B88
				      exchange and BP86 correlation potential
				      and dotted curve is obtained with PW91
				      exchange and correlation potential.

\item [Figure \ref{Fcc_multipol}.] Potential energy curve of CO on Pt(111) at 
                                   the  {\it fcc} hollow position. The Platinum
				   cluster consists of 12 atoms: Pt$_{12}(6,6)$.
				   Solid curve is obtained with B88
				   exchange and BP86 correlation potential
				   and dotted curve is obtained with PW91
				   exchange and correlation potential.
\item [Figure \ref{Hcp_multipol}.]  Potential energy curve of CO on Pt(111) at 
                                    the {\it hcp} hollow position. The platinum 
				    cluster consists of 13 atoms: 
				    Pt$_{13}(6,7)$.
				   Solid curve is obtained with B88
				   exchange and BP86 correlation potential
				   and dotted curve is obtained with PW91
				   exchange and correlation potential.

 \end{description}

\begin{table}[t]
\begin{center}

\newpage
\begin{tabular}[t]{|r|c|c|r|} \hline 
Funktional & RLDA   &  B88/P86 &  Experiment\\ \hline
BE (eV)    & 14.18 & 13.25       & 11.09 \cite{Huber}\\
d$_{C-O}$ (\AA) & 1.148 & 1.153 & 1.128 \cite{Huber}\\ 
IP (eV)    & 15.01 & 15.11       & 14.01 \cite{Huber}\\
$\omega_{C-O} (cm^{-1})$ & 2281.11& 2176.22 & 2169.81 \cite{Huber}\\
$\mu$ (a.u.) & -0.027 & -0.020 & -0.046 \cite{CO-Dipol} \\ \hline
\end{tabular}
\end{center}
\caption{} \label{Multipoltabelle}
\end{table}

\vspace{\fill}

\begin{table}[t]
\begin{center}
\begin{tabular}[t]{|l|c|c|c|c|}  \hline
Energy &  top-    & bridge- & {\it hcp} hollow- & {\it fcc} hollow- \\ 
functional& position & position & position         & position \\ \hline  
RLDA    & 3.25 & 3.15  & 3.35 & 3.25 \\ 
{\bf B88/P86} &{\bf 2.30} & {\bf 1.87}  & {\bf 1.65} & {\bf 1.76} \\ 
Experiment \cite{Yeo} &$1.86 \pm 0.20 $& --- &---&--- \\ \hline 
\end{tabular}
\end{center}
\caption{}\label{Tabelle_mult1}
\end{table}

\vspace{\fill}

\newpage
\begin{table}[t]
\begin{center}
\begin{tabular}[t]{|l|c|c|c|c|}  \hline
Energy &  top-    & bridge- & {\it hcp} hollow- & {\it fcc} hollow- \\
functional& position & position & position         & position \\ \hline
RLDA    & 3.50& 2.60  & 2.55 & 2.55  \\
{\bf B88/P86} & {\bf 3.57} & {\bf 2.70} & {\bf 2.65} & {\bf 2.65}   \\
\hline 
\end{tabular}
\end{center}
\caption{}\label{Tabelle_mult2}
\end{table}

\vspace{\fill}

\newpage
\begin{table}[t]
\begin{center}
\begin{tabular}[t]{|l|c|c|c|c|}  \hline
Energy &  top-    & bridge- & {\it hcp} hollow- & {\it fcc} hollow- \\
functional& position & position & position         & position \\ \hline
RLDA      & 3.50 
          & 3.69 
	  & 3.95 
	  & 3.95 
	  \\
{\bf B88/P86} &{\bf 3.57} 
             &{\bf 3.76} 
	     &{\bf 4.03} 
	     &{\bf 4.03} 
	      \\
\hline
Experiment \cite{Ogletree,Blackman} 
            &$3.50\pm 0.19$ 
	    &$3.93\pm 0.13$ 
	    &---&---
\\ \hline
\end{tabular}
\end{center}
\caption{}\label{Vergleich-Abstand}
\end{table}

\vspace{\fill}

\newpage

\begin{table}[t]
\begin{center}
\begin{tabular}[t]{|l|c|c|c|c|}  \hline
method  & r$_{Pt-C} [a.u.]$  & E$_b$(eV) & Ref. & Explanation \\ \hline
RLDA & 3.50 & 3.25 & this work & 13 Pt atoms, 2 layers (7,6) \\
RGGA & 3.57 & 2.30 & this work & 13 Pt atoms, 2 layers (7,6) \\
CASSCF &  3.84      &  1.44 & \cite{Bala3} & Complete Active Space \\
PP-Slab &  3.55 & 1.45 & \cite{Hammer} & 6 layers slab \\
ZORA-DFT GGA &  3.59  & 1.41 & \cite{Philipsen2} & spin-orbit, 2 layers
slab \\
exp.         &  $3.49\pm0.19$   &  & \cite{Ogletree,Blackman} & \\ \hline 
\end{tabular}
\end{center}
\caption{} \label{theory-comparision}
\end{table}

\vspace{\fill}

\newpage
\begin{table}[t]
\begin{center}
\begin{tabular}[t]{|l|c|c|c|l|}  \hline
method  & r$_{Pt-C} [a.u.]$  & E$_b$(eV) & Ref. & Explanation \\ \hline
RLDA & 3.95 & 3.25 & this work & 9 Pt atoms; 2 layers (6,3)  \\
RGGA & 4.02 & 1.76 & this work & 9 Pt atoms; 2 layers (6,3)  \\
ORPP &  3.96      &  2.39     &  \cite{Gavezotti} & optimized relativistic pseudo potential \\
ZORA-DFT GGA &  4.12      &  1.05     & \cite{Philipsen2} & pin-orbit, 2
layers slab \\ 
exp.         &  $3.93\pm0.13$   & & \cite{Ogletree,Blackman} &  \\ \hline
\end{tabular}
\end{center}
\caption{} \label{theory-comparision1}
\end{table}

\vspace{\fill}

\newpage
\begin{table}[t]
\begin{center}
\begin{tabular}[t]{|l|c|c|c|c|c|}   \hline
          &  top-    & bridge- & {\it hcp} hollow- & {\it fcc} hollow-  & free\\
          & position & position & position& position   & molecule\\ \hline
 Oxygen  & 8.375  & 8.370 & 8.359 &  8.359  & 8.348\\
 Carbon& 5.719  & 6.072 & 6.094 &  6.075  & 5.652\\
 Sum    &  14.094 & 14.442 & 14.453 & 14.434 & 14.000  \\ \hline
\end{tabular}
\end{center}
\caption{}\label{Mulliken_CO}
\end{table}

\newpage

\vspace{\fill}

\vspace{\fill}

\vspace{\fill}

\end{document}